\documentclass[12pt,preprint]{emulateapj}

\shorttitle{Trapezium Cluster Disk Mass Distribution}
\shortauthors{Mann \& Williams}

\received{2008 Nov 26}
\slugcomment{{\sc Accepted to ApJL:} 2009 Feb 3}
\begin{document}

\title{The Circumstellar Disk Mass Distribution in the Orion Trapezium Cluster}
\author{Rita K. Mann, \and Jonathan P. Williams}

\affil{Institute for Astronomy, University of Hawaii, 2680 Woodlawn Drive, Honolulu, HI 96822}
\email{rmann@ifa.hawaii.edu, jpw@ifa.hawaii.edu}

\begin{abstract}
We present the results of a submillimeter interferometric survey of
circumstellar disks in the Trapezium Cluster of Orion.
We observed the $880\,\mu$m continuum emission from 55 disks
using the Submillimeter Array\footnote{
The Submillimeter Array is a joint project between the
Submillimeter Astrophysical Observatory and the Academica Sinica Institute
of Astronomy and Astrophysics and is funded by the Smithsonian Institution
and the Academica Sinica.
}, and detected 28 disks above $3\sigma$
significance with fluxes between 6-70 mJy and rms noise between 0.7--5.3 mJy.
Dust masses and upper limits are derived from the submillimeter excess
above free-free emission extrapolated from longer wavelength observations.
Above our completeness limit of $0.0084\,M_\odot$,
the disk mass distribution is similar to that of Class II disks in
Taurus-Auriga and $\rho$\,Ophiuchus but is truncated at$~0.04\,M_\odot$.
We show that the disk mass and radius distributions are consistent with
the formation of Trapezium Cluster disks $\sim 1$\,Myr ago and subsequent
photoevaporation by the ultraviolet radiation field from $\theta^1$\,Ori\,C.
The fraction of disks which contain a minimum mass solar nebula within
60\,AU radius is estimated to be $11-13$\% in both Taurus and 
the Trapezium Cluster, which suggests the potential for forming 
Solar Systems is not compromised in this massive star-forming region.
\end{abstract}


\section{Introduction}\label{sec: intro}
Circumstellar disks are the birth sites of planets, making their fundamental
properties, such as mass and size, directly related to the types of planets
that can form.  Disk masses are most easily measured by observations at
millimeter wavelengths, where the dust grain optical depth is less than unity. 
Surveys of stars with ages $\sim 0.3-10$\,Myr in the Taurus and $\rho$\,Ophiuchus 
dark clouds show a broad range of disk masses around a median of 
$0.005\,M_\odot$ \citep{andrews05,andrews07}, which is within a factor of two 
required to form our Solar System (the minimum mass solar nebula; 
MMSN$=0.01\,M_\odot$ \citep{weidenschilling}.
Most studies to date of disk masses have focused on Taurus and $\rho$\,Ophiuchus
because they are the nearest star-forming regions. However, most stars
do not form in relative isolation as in these regions, but in dense
clusters containing high mass stars \citep{lada}. The Orion Trapezium Cluster 
is more representative of the typical birthplace of most stars,
including our Sun.  There are more than 2000 stars with mean ages less than
1\,Myr packed within 1\,pc$^{3}$ in the Trapezium \citep{hillenbrand97}.
Hubble Space Telescope (HST) images of this region provided the first
direct images of circumstellar disks (dubbed ``proplyds''),
spectacularly seen silhouetted against the bright background \citep{odell94,odell96,bally00}.
Cometary-shaped cocoons of ionized gas surround many disks, excited 
by ultraviolet radiation from the most massive and luminous (O6)
star at the center of the cluster, $\theta^1$\,Ori\,C \citep{mccullough,bally98a}.
Centimeter wavelength observations have revealed photoevaporative mass
loss rates of $\dot M\sim 10^{-7}\,M_\odot\,{\rm yr}^{-1}$,
which are sufficient to remove a MMSN outside the gravitationally
bound radius in $\sim$1\,Myr \citep{churchwell,henney}.

Mass loss is expected to be concentrated in the outer parts of
the disk with the inner regions potentially surviving longer 
\citep{johnstone98,adams,clarke}.  But the amount of material 
in the inner disk regions available to form planets was unknown.
Weak lower limits on disk mass
exist from the extinction of the central star and background
nebula \citep{mccaughrean} but measurements of optically thin disk emission
are required for comparison with Taurus and other star-forming regions.
Detecting dust emission from disks in the Trapezium Cluster is
challenging because of its distance (3 times further than Taurus),
the small angular separation between disks, and substantial levels
of ionized gas from the surrounding cocoons, which swamp dust
emission at centimeter to millimeter wavelengths \citep{mundy}.
The Submillimeter Array (SMA) is ideally suited to detecting
thermal dust emission from the Trapezium Cluster disks,
as it overcomes the above problems with its combination of
high sensitivity, high resolution, and high frequency.

We have carried out a survey of HST-identified proplyds in 
the Trapezium Cluster using the SMA in order to determine the disk
mass distribution. We describe the observations in \S2,
the determination of masses and comparison with Taurus
and $\rho$\,Ophiuchus in \S3, and discuss the implications
of our findings for disk evolution and the possibility of
planet formation in \S4.

\section{Observations}
Submillimeter interferometric observations at $880\,\mu$m  
were conducted with the SMA on Mauna Kea over 9 nights between 2006-08 
(see Figure~1\footnote{HST image from http://casa.colorado.edu/$\sim$bally/HST/HST/master/.}).
The compact configuration of the 8-element
interferometer was chosen to provide the best phase stability,
and sufficient resolution ($\sim 2.5''$) to distinguish individual disks.
With the exception of the central region around $\theta^1$\,Ori\,C,
fields were chosen to lie in regions of relatively low and
uniform background cloud emission.
Field B was intentionally chosen to overlap the 2004 observations
of \citet{williams} because of improvements in array sensitivity.

Observations were carried out during good weather
conditions with precipitable water vapor levels below 2\,mm,
resulting in system temperatures ranging from 100-400\,K.
Amplitude and phase calibration were performed through
observations of the quasars J0423-013 and J0530+135.
Passband calibration was conducted with bright quasars 3C454.3,
3C279, or 3C273 and Uranus or Titan were used to set the absolute
flux scale, which is estimated to be accurate to $\sim$ 10\%.

The Trapezium Cluster has formed a blister HII region and sits
in front of a Giant Molecular Cloud.
Bolometer maps made with the SCUBA camera on the James Clerk Maxwell
Telescope by \citet{johnstone99} show that the cloud produces a strong $850\,\mu$m
background of several Jy per $15''$ beam. This corresponds to $\sim 10$\,mJy
per square arcsecond and is comparable to disk emission.
We produced images using physical baselines greater than
23\,m (27k$\lambda$) to filter extended emission on angular
scales $\gtrsim 7''$, which lowered the cloud background substantially
but preserved compact emission from the disks.
Nevertheless, the background contributed significantly to the
noise level in several fields.

\begin{figure}[ht]
\epsscale{1.35}
\plotone{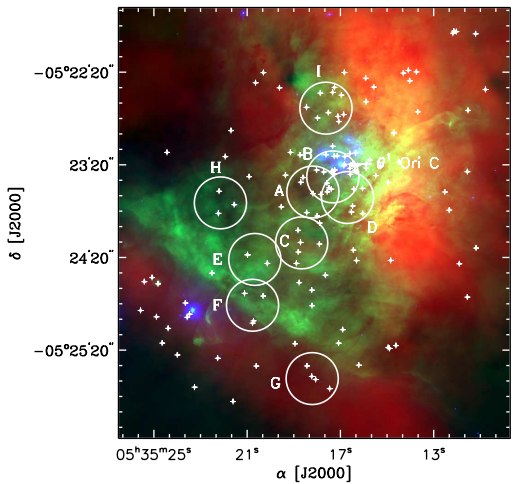}
\caption{Large-scale view of the Orion Trapezium Cluster as a color
mosaic of images from the Hubble Space Telescope (HST) and James Clerk
Maxwell Telescope (JCMT).  JCMT/SCUBA $450\,\mu$m observations are
in red (from \citet{johnstone99}), HST F555W in blue and H$\alpha$ in
green$^2$.  Blue stars near the center of the image are the four OB stars
that give the Trapezium its name.  The most massive member, $\theta^1$\,Ori\,C,
is labelled.  Solid circles represent the $32''$ primary beam of each SMA field.}
\end{figure}

\begin{figure}[ht]
\epsscale{1.0}
\plotone{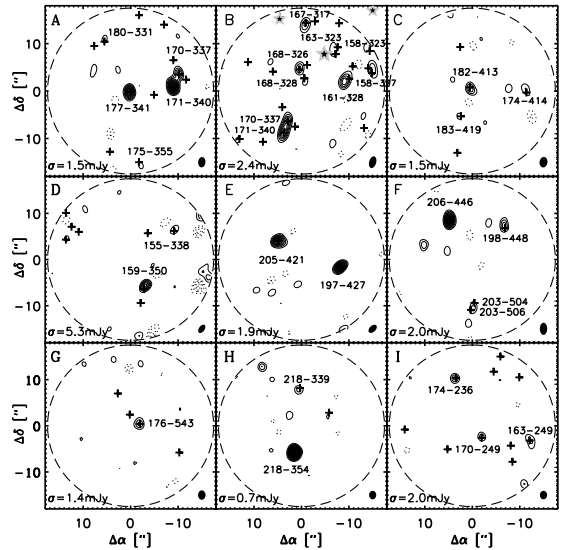}
\vskip -0.6in
\caption{Continuum emission at $880\,\mu$m toward 9 SMA fields.
Alphabetical labels correspond to pointings in Figure~1.
The 55 HST-identified disks that lie within the $32''$ SMA
primary beams (shown as dashed circles) are marked with crosses
and the 28 detections are labeled using standard proplyd nomenclature.
The OB stars are shown with star symbols in Field B.
The contours in each field begin at $3\sigma$ and increase
by $2\sigma$, where the rms noise level, $\sigma$, is specified in the
lower left corner. The synthesized beam
is shown in the lower right corner of each map.}
\end{figure}

\bigskip
\section{Results}
\subsection{Disk mass determination}
The 9 SMA fields include a total of 55 HST-identified disks from the
catalogs of \citet{odell96,bally00,vicente} and are shown in Figure 2.
We detected 28 disks with a signal-to-noise ratio
$\geq 3$ and refer to them in the standard proplyd nomenclature \citep{odell94}.
Given the relatively large beam size ($\sim 1000$\,AU), all the detections
are point sources and we measured their fluxes in a beam-sized aperture
centered on the HST position.
The observed submillimeter fluxes are composed of 3 components:
blackbody emission from the dust in the disks, F$_{\rm dust}$,
free-free emission from the cocoons of ionized gas, F$_{\rm ff}$,
and background cloud emission, F$_{\rm bg}$:
\begin{equation} \label{equ: fluxes}
F_{\rm obs} = F_{\rm dust} + F_{\rm ff} + F_{\rm bg}.
\end{equation}

The radio-millimeter spectral energy distributions (SED)
for each SMA detected disk are shown in Figure~3.
The free-free emission from several disks detected at long
wavelengths was extrapolated into the submillimeter regime
using published fluxes from 6\,cm to 1.3\,mm 
\citep{churchwell,garay,felli,zapata,mundy,bally98b,eisner06,eisner08}.
These data show a flat spectral dependence consistent with
optically thin emission, $F_\nu\propto\nu^{0.1}$, but with a range,
shown as grey scale, which we attribute to variability \citep{zapata}
and possibly different amounts of background filtering.
The turnover to optically thick emission occurs at longer
wavelengths than shown and we have not attempted to model this.
All but two $880\,\mu$m SMA fluxes exceed the extrapolated
free-free emission indicating disk or background emission.
A template disk spectrum, $F_\nu\propto\nu^2$ \citep{andrews05},
is schematically plotted through the SMA data in Figure~3 to guide the eye.

The background at $880\,\mu$m was estimated by simulating the
interferometric response to the \citet{johnstone99} SCUBA
map. For each field, the SCUBA data were Fourier-transformed and
sampled over the same $uv-$tracks as the observation, then
inverted and cleaned to produce a spatially filtered map.
The background flux at each disk position was then measured in the
same manner as the observations themselves.

The dust-disk flux was determined by subtracting
free-free and background contributions from the observed flux (Table~1).
We found that 26 of our 28 detections had significant dust emission.
Disk masses, and upper limits for the non-detections, 
were derived based on the standard relationship,
\begin{equation} \label{equ: masses}
M_{\rm disk} = \frac{F_{\rm dust}d^2}{\kappa_{\nu}B_{\nu}(T)},
\end{equation}
where $d=400$\,pc is the distance to Orion \citep{sandstrom,menten},
$\kappa_{\nu}=0.1(\nu/1000\,{\rm GHz})=0.034\,{\rm cm}^2\,{\rm g}^{-1}$
is the dust grain opacity with an implicit gas-to-dust mass ratio of 100:1,
and B$_{\nu}(T)$ is the Planck function at temperature $T=20$\,K.
We used the same dust opacity and temperature as the
disk surveys of Taurus and $\rho$\,Ophiuchus by \citet{andrews05,andrews07}
since we wish to make a direct comparison with these studies,
but revisit these assumptions in \S4.

We also derived the mass sensitivity of the survey by determing
the fraction of sources that could be detected at $\geq 3\sigma$
at each mass.  We input disk masses and observed free-free emission,
adding appropriate point sources to the SCUBA map, and simulating
observations in the same manner as described above for the background
contamination.
Our survey is 100\% complete for disk masses $\geq 0.0084\,M_\odot$
and 85\% complete for masses a factor of two lower.

\begin{figure}[ht]
\epsscale{1.0}
\plotone{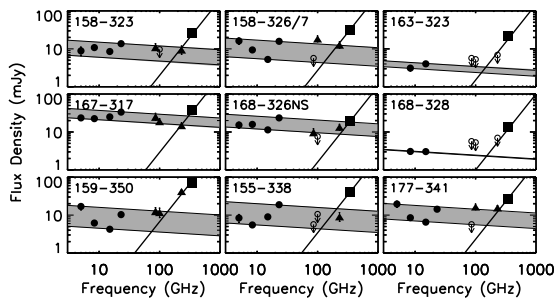}
\vskip -2.1in
\caption{Spectral energy distributions for 9 proplyds detected at
$\geq 3\sigma$ with the SMA at $880\,\mu$m.
The SMA measurements are represented by squares, millimeter
observations by triangles and centimeter observations by
circles.  Open circles are upper limits from non-detections and
uncertainties not shown are smaller than symbol sizes.
The extrapolated range of optically thin free-free emission,
$F_\nu \propto \nu^{-0.1}$, is overlaid in gray.
A template to the disk emission, $F_\nu \propto \nu^2$,
shows the relative contributions of the ionized gas and dust components.}
\end{figure}

\begin{deluxetable*}{lccccccc}
\tablecolumns{8}
%\tablewidth{4in}
\tabletypesize{\scriptsize}
\tablecaption{Disk fluxes and masses \label{}}
\tablehead{
\colhead{Proplyd} & \colhead{Field} & \colhead{$F_{\rm obs}$} & \colhead{rms} &
\colhead{$F_{\rm ff}$} & \colhead{$F_{\rm bg}$} & \colhead{$F_{\rm dust}$} &
\colhead{$M_{\rm disk}$} \\
\colhead{Name} & \colhead{} & \colhead{(mJy)} & \colhead{(mJy)} &
\colhead{(mJy)} & \colhead{(mJy)} & \colhead{(mJy)} &
\colhead{($10^{-2}\,M_\odot$)} \\
\colhead{(a)} & \colhead{(b)} & \colhead{(c)} & \colhead{(d)} &
\colhead{(e)} & \colhead{(f)} & \colhead{(g)} & \colhead{(h)}
}
\startdata 
155-338  & D & 39.6 & 5.3 & 4.0--14.0 &  2.6  &  23.0  & 1.13 $\pm$ 0.34  \\
158-323  & B & 25.5 & 2.4 & 4.3--11.5 & -5.6  &  19.5  & 0.96 $\pm$ 0.21 \\
158-326/7 & B & 31.9 & 2.4 & 4.0--12.0 & -0.5  &  20.3  & 1.00 $\pm$ 0.19 \\
159-350  & D & 69.8 & 5.3 & 3.2--11.0 &  0.3  &  58.4  & 2.86 $\pm$ 0.29 \\
161-328   & B & 38.7 & 2.4 & 0.1--1.5  & -1.2  &  38.4  & 1.88 $\pm$ 0.15 \\
163-249   & I & 34.8 & 2.0 &   0.0     & -1.6  &  36.4  & 1.78 $\pm$ 0.13 \\
163-323   & B & 22.1 & 2.4 & 2.4--3.0  &  0.2  &  18.9  & 0.93 $\pm$ 0.15 \\
167-317   & B & 37.2 & 2.4 & 14.5--25.5& -5.4 &  17.0  & 0.83 $\pm$ 0.18 \\
168-328   & B & 13.7 & 2.4 & 1.9--2.5  &  1.0 &  10.3  & 0.50 $\pm$ 0.12 \\
170-249   & I & 14.6 & 2.0 & 0.0--2.5  & -0.6 &  12.6  & 0.62 $\pm$ 0.10 \\
170-337   & A & 19.1 & 1.5 & 4.0--9.0  &  3.0 &   7.1  & 0.32 $\pm$ 0.08 \\
171-340   & A & 46.4 & 1.5 &    0.0    & -2.8 &  49.2  & 2.23 $\pm$ 0.08 \\
174-236   & I & 25.2 & 2.0 &    0.0    &  2.9 &  22.2  & 1.09 $\pm$ 0.13 \\
174-414   & C & 12.7 & 1.5 &    0.0    &  1.6 &  11.1  & 0.51 $\pm$ 0.09 \\
175-355   & A &  9.4  & 1.5 &    0.0    & -0.1 &   9.5  & 0.43 $\pm$ 0.11 \\
176-543   & G & 15.8  & 1.7 &    0.0    & -0.6 &  13.4  & 0.89 $\pm$ 0.09 \\
177-341   & A & 26.9  & 1.5 & 5.0--13.0 & -4.9 &  19.8  & 0.90 $\pm$ 0.07 \\
182-413   & C & 13.3  & 1.5 &    0.0    & -3.0 &  16.4  & 0.74 $\pm$ 0.07 \\
183-419   & C &  6.2  & 1.5 &    0.0    &  0.2 &   6.1  & 0.28 $\pm$ 0.07 \\
197-427   & E & 57.7  & 1.9 &    0.0    & -1.5 &  59.3  & 2.72 $\pm$ 0.10 \\
198-448   & F & 16.4  & 2.0 &    0.0    & -2.9 &  19.3  & 0.94 $\pm$ 0.12 \\
203-506/4 & F & 14.3  & 2.0 &    0.0    & -2.9 &  17.2  & 0.84 $\pm$ 0.12 \\
205-421   & E & 54.8  & 1.9 &    0.0    & -1.7 &  56.5  & 2.59 $\pm$ 0.10 \\
206-446   & F & 63.3  & 2.0 &    0.0    &  0.0 &  63.3  & 3.10 $\pm$ 0.12 \\
218-339   & H &  6.9  & 0.9 &    0.0    &  0.0 &   6.6  & 0.34 $\pm$ 0.04 \\
218-354   & H & 42.9  & 0.9 &    0.0    & -0.9 &  48.1  & 2.10 $\pm$ 0.04 \\
\enddata 
\tablecomments{
(a) Proplyd designation based on the nomenclature of O'Dell \& Wen (1994).
(b) Observed Field in Figure~1.
(c) Integrated continuum flux density, corrected for SMA primary beam attenuation.
(d) 1$\sigma$ statistical error.
(e) Range of extrapolated contribution of free--free emission at $880 \,\mu$m.
(f) Estimated flux contribution from cloud background. 
(g) Derived dust continuum flux from the disk.
(h) Disk mass.
}
\end{deluxetable*}

\subsection{Comparison with Taurus and $\rho$\,Ophiuchus}
Having measured masses for a moderate sample of Trapezium Cluster 
disks, we can address how they compare with disks in 
Taurus and $\rho$\,Ophiuchus and assess the effect of
environment on disk evolution. 
We use the results of the SCUBA surveys of \citet{andrews05,andrews07} 
and consider only Class II disks as they have ages $\sim 1$\,Myr,
similar to those estimated for the Trapezium Cluster.
The differential disk mass distributions for the three samples
are plotted in Figure~4 with the same bins, $\Delta\log\,M_{\rm d}=2$,
and with a start point that separates the complete and incomplete
samples in the Trapezium Cluster.

\begin{figure}[ht]
\epsscale{1.0} 
\plotone{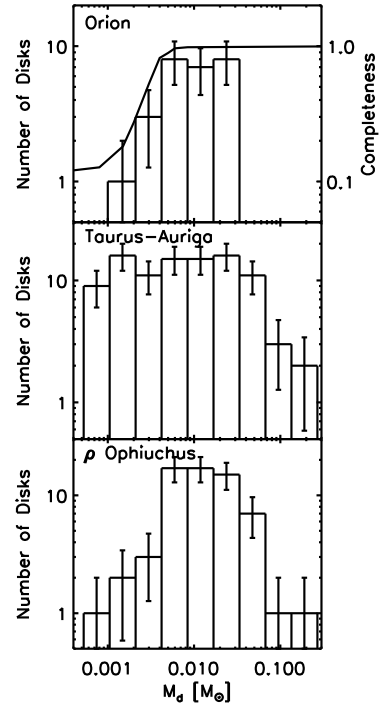}
%\vskip -0.2in
\caption{Differential disk mass distribution in Orion,
Taurus and $\rho$\,Ophiuchus.  The error bars for the distributions
are $\sqrt{N}$.  The fraction of disks that would be detected in our
survey at each mass is indicated by the continuous line in the top panel.
Binning begins at $0.0084\,M_\odot$ to separate complete and 
incomplete samples. The number of disks per logarithmic mass bin is
approximately constant for masses $0.004-0.034\,M_\odot$
in all three regions but there are no disks with masses
$> 0.034\,M_\odot$ in Orion.}
\end{figure}

Figure~4 shows that the number of disks per logarithmic mass is
approximately constant for the 3 bins spanning $0.004-0.034\,M_\odot$
in each region but there are no Trapezium disks in the
higher mass bins. This cannot be explained by observational biases,
since such massive disks would be readily detectable in our observations.
If the Trapezium disk mass distribution were similar to
Taurus (or $\rho$\,Ophiuchus), we would expect 8 (4) disks with
masses $>0.034\,M_\odot$ in our sample.  With this expectation, 
the probability of detecting none is 0.03\% (1.8\%) respectively.
Two-tailed Kolmogorov-Smirnov (KS) tests on the disk distributions
above $0.0084\,M_\odot$ for which all observations are fully complete,
show that the Trapezium disk distribution is different from that
in both Taurus and $\rho$\,Ophiuchus at $>5\sigma$ significance.
The difference remains significant if the disk luminosities
are compared, and shows that the flux-mass conversion is not responsible
for the observed discrepancy.  We conclude that the disk mass distribution 
in the Orion Trapezium Cluster is truncated at high masses.

\citet{eisner08} also noted a lack of massive disks in the
Trapezium Cluster from a larger scale but longer wavelength survey.
We agree with their finding but our results differ in the details.
First, we generally detect significant $880\,\mu$m emission only
toward HST-identified disks. There are just two points in our
9 maps that exceed $5\sigma$ and are not coincident with a
catalogued proplyd. In particular, we do not detect other known objects
such as the infrared sources cataloged by \citet{hillenbrand00}.
We suspect that at least some of the HC and MM sources in
\citet{eisner08} are filtered small-scale background fluctuations
despite their precaution in using localized measures of the noise.
Second, the SEDs in Figure~3 show that several of the detections
listed at 3\,mm by \citet{eisner06} and 1.3\,mm by \citet{eisner08}
are consistent with being free-free, rather than disk, emission.
Although they attempted to separate contributions
from dust and free-free emission, this process is much
easier at shorter wavelengths.  Further, we were more conservative,
using a range to allow for variability in the radio emission.
The difficulty in separating out the contributions from the ionized
gas and the dust and, to a lesser extent, the cloud background,
results in several significant discrepancies in individual disk mass
determinations between our work and theirs.

\section{Discussion}
We first address the flux-to-mass conversion in equation~\ref{equ: masses}.
Our prescription for the dust grain opacity, $\kappa$ is from \citet{beckwith}
but its value depends on the grain composition and size distribution.
Detailed models by \citet{pollack} and \citet{ossenkopf} show that, excluding 
the diffuse ISM, the range in $\kappa(1\,{\rm mm})$ is about a factor of 4.
A low surface area to mass is required to explain the observed low
maximum disk luminosity in the Trapezium Cluster.  \citet{throop} have 
proposed that gas loss through photoevaporation promotes dust coagulation 
but since the former mainly occurs in the outer disk, it does not appear 
to be applicable to the many small disks detected here. However, given that 
there are several indications of significant grain growth in Taurus and 
$\rho$\,Ophiuchus Class II disks \citep{andrews05,andrews07}, there is 
insufficient remaining leverage in $\kappa$ to account for the lower disk 
luminosities in the Trapezium Cluster.  Additional evidence against a large 
range in $\kappa$ is the similarity of the Taurus and $\rho$\,Ophiuchus
disk mass distributions \citep{andrews07}.  We have assumed a low 
temperature, 20\,K, derived from the average for Taurus disks
\citep{andrews05} however, the high radiation field might significantly heat 
the Orion disks.  But a higher temperature would reduce the inferred 
Trapezium Cluster disk masses yet further.  The only remaining variable is the distance to 
the Trapezium, which has been securely determined from recent VLBI parallax 
observations \citep{menten,sandstrom} and the small error cannot account 
for the luminosity difference.  Finally, we note that different scalings 
do not change the shape of the distributions and our finding that the 
Trapezium Cluster disk mass distribution is truncated at its upper end.

The lack of massive disks in the Trapezium Cluster is in
good agreement with theoretical models of disk photoevaporation
\citep{adams,johnstone98}.
Photoevaporation is thought to be the dominant external influence on
disk evolution in the Trapezium Cluster as the rate of disk-disk 
encounters is low \citep{scally}.  The models predict that mass loss is
highest for the largest disks because material at large radii is unbound 
to the embedded star and provides a greater surface area
for photoevaporation.  In contrast, the inner regions of the disk are
gravitationally bound and can survive longer: while
it takes only 2\,Myr to erode disks down to 50\,AU size
scales, the dust and gas at smaller radii may survive external
photoevaporation for $>$\,10\,Myr \citep{adams,clarke}.
Consequently, the largest disks, which are likely to be the most massive,
would be quickly eroded to smaller radii and lower masses. The factor of
$\sim 3$ reduction between the upper end of the Trapezium Cluster
versus the Taurus and $\rho$\,Ophiuchus disk mass distributions
corresponds to a reduction of $3-9$ in radius for a power law surface
density profile, $\Sigma(r) \sim r^{-p}$, with $p=1-1.5$
\citep{andrews07b}. If the Trapezium Cluster disks started out
with initial properties similar to Taurus disks, which have a mean radii
of 200\,AU, range 100--700 AU and masses $0.01-0.17\,M_\odot$
\citep{andrews07b}, we would expect them to be eroded to scales $<70$\,AU
and masses consistent with our survey.  In fact, many of the
Trapezium Cluster disks in our sample are unresolved by HST
observations implying radii $<60$\,AU.
And in the cases where the proplyds are seen in silhouette
and the radius can be directly measured \citep{vicente},
we find that the largest disks, R$\sim 150$\,AU,
tend to be among the most massive, $M_{\rm d}\sim 0.03\,M_\odot$.  

The truncation of the Trapezium Cluster disk mass distribution is
a vivid illustration of the the effect of nearby massive stars on
disk evolution.  The hostile environment, which includes both 
ionizing radiation and high density of stars, may preclude the
formation of massive disks in their vicinity, or rapidly erode them
after formation.
Yet the effect of external photoevaporation on the inner, potentially 
planet-forming, regions of disks appears to be small.
12 of the disks in our sample exceed the MMSN and retain
sufficient material to form planetary systems with masses like our own.
Half of these (11\% of the total sample) are unresolved with radii
$<60$\,AU making them resistant to further photoevaporation. 
For comparison, $\sim$30\% of Taurus and $\rho$\,Ophiuchus
Class II disks exceed a MMSN but these are generally larger
than the Trapezium Cluster disks. A more meaningful comparison would be to
estimate the fraction of disks containing a MMSN within their central 60\,AU.
This would imply a total mass out to 200\,AU of
$\gtrsim 3$\,MMSN, which is only found in about 13\% of Taurus disks.
The fraction of disks that appear capable of forming Solar Systems
is similar in each region and lies between the $\sim 7$\% detection rate
of extrasolar planets in Doppler surveys \citep{marcy} 
and the $\sim 15$\% debris disk fraction \citep{carpenter}.

The formation of sub-Jupiter mass planets requires less
material and may be possible for proplyds less massive than 
a MMSN.  For example, 11 proplyds have sufficient 
mass ($\sim 1/3$ MMSN) and small sizes ($<60$\,AU) 
to potentially form Neptune.  This fraction (20\%) is a lower limit as our
survey completeness drops rapidly for $M_{\rm d}< 0.004\,M_\odot$,
but is on the lower end of a recent first estimate from the HARPS
planet-search survey of 30\% $\pm$ 10\% \citep{mayor}.

\section{Summary}
Our SMA survey of the Orion Trapezium Cluster detected 28 out of 55
HST-identified disks (``proplyds'') at $880\,\mu$m 
due to its combination of high sensitivity and resolution.
We carried out a careful analysis of the disk SED
from centimeter to submillimeter wavelengths
and of the interferometric response to the cloud background
to calculate the dust flux from each disk.
We then determined dust masses for 26 disks and show the
number of disks per logarithmic mass interval is approximately
constant over almost a decade in mass between $0.004-0.034\,M_\odot$,
similar to Class II disks in both Taurus and $\rho$\,Ophiuchus.
Extrapolating from these low mass star forming regions,
we would have expected several disks with masses greater
than the maximum in our survey, $0.034\,M_\odot$.
The truncation of the Trapezium Cluster disk mass distribution
is likely due to photoevaporation of the outer disks by the 
O6 star, $\theta^1$\,Ori\,C, from an initial disk distribution
similar to Taurus.

Six of the 55 disks exceed the MMSN and have radii $<60$\,AU,
which is similar to the inferred initial conditions of the Solar System,
whose sharp edge at $\sim 50$\,AU may also be a product of
photoevaporation \citep{jewitt,trujillo,allen}.
We conclude that disks with the potential to form Solar Systems
are no less likely in Orion than in dark clouds without
massive stars.

\acknowledgments
This work is supported by the NSF through grant AST06-07710.
We are grateful to David Jewitt, Michael Liu, Sean Andrews,
and David Wilner for their comments.

\end{document}